\documentclass[%
 reprint,
 amsmath,amssymb,
prl,
]{revtex4-2}

\usepackage{lipsum}
\usepackage{bm}
\usepackage{dblfloatfix}
\usepackage{amssymb}
\usepackage{amstext}
\usepackage{amsmath}
\usepackage{mathrsfs}
\usepackage{graphicx}
\usepackage{color}
\usepackage{amsthm}
\usepackage{floatrow}
\usepackage{subfig}
\usepackage{array}

\usepackage{listings}
\usepackage{booktabs}
\usepackage{hyperref}

\begin{document}
\preprint{APS/123-QED}
\title{A unified theory of information transfer and causal relation}
\thanks{Correspondence of this research should be addressed to Y.T. and P.S.}%
\author{Yang Tian}
\email{tiany20@mails.tsinghua.edu.cn}
 \altaffiliation[]{Department of Psychology \& Tsinghua Laboratory of Brain and Intelligence, Tsinghua University, Beijing, 100084, China.}
  \altaffiliation[Also at ]{Laboratory of Advanced Computing and Storage, Central Research Institute, 2012 Laboratories, Huawei Technologies Co. Ltd., Beijing, 100084, China.}
  \author{Hedong Hou}
\email{hedong.hou@etu.u-paris.fr}
 \altaffiliation[]{UFR de Math\'{e}matiques, Universit\'{e} de Paris, 75013 Paris, France.}
  \author{Yaoyuan Wang}
\email{wangyaoyuan1@huawei.com}
 \altaffiliation[]{Laboratory of Advanced Computing and Storage, Central Research Institute, 2012 Laboratories, Huawei Technologies Co. Ltd., Beijing, 100084, China.}
  \author{Ziyang Zhang}
\email{zhangziyang11@huawei.com}
 \altaffiliation[]{Laboratory of Advanced Computing and Storage, Central Research Institute, 2012 Laboratories, Huawei Technologies Co. Ltd., Beijing, 100084, China.}
\author{Pei Sun}%
 \email{peisun@tsinghua.edu.cn}
 \altaffiliation[]{Department of Psychology \& Tsinghua Brain and Intelligence Lab, Tsinghua University, Beijing, 100084, China.}



\begin{abstract}
Information transfer between coupled stochastic dynamics, measured by transfer entropy and information flow, is suggested as a physical process underlying the causal relation of systems. While information transfer analysis has booming applications in both science and engineering fields, critical mysteries about its foundations remain unsolved. Fundamental yet difficult questions concern how information transfer and causal relation originate, what they depend on, how they differ from each other, and if they are created by a unified and general quantity. These questions essentially determine the validity of causal relation measurement via information transfer. Here we pursue to lay a complete theoretical basis of information transfer and causal relation. Beyond the well-known relations between these concepts that conditionally hold, we demonstrate that information transfer and causal relation universally originate from specific information synergy and redundancy phenomena characterized by high-order mutual information. More importantly, our theory analytically explains the mechanisms for information transfer and causal relation to originate, vanish, and differ from each other. Moreover, our theory naturally defines the effect sizes of information transfer and causal relation based on high-dimensional coupling events. These results may provide a unified view of information, synergy, and causal relation to bridge Pearl’s causal inference theory in computer science and information transfer analysis in physics.

\end{abstract}

\maketitle
When multiple coupled systems co-evolve across time, \emph{information transfer}, where systems exchange information with each other, frequently occurs \cite{schreiber2000measuring}. Information transfer suggests latent causal relation between systems \cite{schreiber2000measuring} and contributes to entropy production \cite{prokopenko2014transfer}, being critical for understanding diverse complex system behaviours in physics (e.g., emergent traveling coherence \cite{lizier2008local} and chaotic communication \cite{hung2008chaotic}), computer science (e.g., thermodynamic properties of computation \cite{prokopenko2014transfer}), biology (e.g., neural information propagation \cite{wibral2011transfer,shih2015connectomics}, gene regulatory mechanisms \cite{kim2021tenet}, and biochemical reactions \cite{imaizumi2022assessing}), and social sciences (e.g., information-driven collective behaviours \cite{borge2016dynamics} and stock market dynamics \cite{yao2020effective}). As various information transfer phenomena are discovered across different disciplines, the measurement of transferred information quantities become increasingly significant. \emph{Transfer entropy} \cite{schreiber2000measuring} and \emph{information flow} \cite{ay2008information} are successively proposed to achieve this objective. Although transfer entropy is defined to measure conditional correlations in causal channels \cite{schreiber2000measuring} and information flow is developed based on Pearl’s probabilistic formulation of causal inference \cite{ay2008information}, these two metrics are intrinsically related to conditional mutual information \cite{schreiber2000measuring,ay2008information,lizier2010differentiating}. In past decades, considerable efforts have been devoted to propose appropriate estimation of these metrics (e.g., see Refs. \cite{tian2021fourier,staniek2008symbolic,lobier2014phase,wollstadt2014efficient,restrepo2020transfer,zhang2019itene,kraskov2004estimating,kugiumtzis2013partial}) to overcome computational deficiencies of information transfer analysis on empirical data, such as dimensionality curse \cite{runge2012escaping,tian2021fourier}, noise sensitivity \cite{smirnov2013spurious}, and restraints by stationary assumption \cite{tian2021fourier}. These contributions lay solid computational foundations of measuring information transfer in practice.

Critical problems, however, arise when one treats information transfer as the causal relation studied by Pearl’s causal inference theory \cite{pearl2000models}. It should be emphasized that \emph{information transfer is related to, rather than equivalent to, causal relation, especially when information transfer is quantified by transfer entropy} \cite{lizier2010differentiating}. One can see a comprehensive analysis of their in-equivalence in Ref. \cite{lizier2010differentiating}. This in-equivalence relation naturally leads to several fundamental yet difficult questions: 
\begin{itemize}
    \item[(I) ] How does information transfer arise and vanish?
    \item[(II) ] How does causal relation arise, vanish, and differ from information transfer?
    \item[(III) ] If information transfer and causal relation are special cases of a unified and general variable?
\end{itemize}

Although questions (I-II) have been partially analyzed in Ref. \cite{lizier2010differentiating}, much unknown still remains for further exploration. In this letter, we discuss a unified answer of questions (I-III) by developing a new analytic theory. 

\emph{Mathematical conventions}.---Because we need to deal with long mathematical expressions in this article, we choose an equivalent yet shorter definition of the expectation of random variable $\mathsf{X}$ in the measure theory \cite{cohn2013measure,ash2000probability}
\begin{align}
\langle \mathsf{X}\rangle=\int_{\Omega}x\mathsf{d}\rho\left(x\right)\equiv\int_{\Omega}\rho\left(x\right)x\mathsf{d}x,\label{EQ1}
\end{align}
where $\rho\left(\cdot\right)$ denotes a probability measure and $\Omega$ is the support set. Note that we can readily derive 
\begin{align}
\int_{\Omega}\rho\left(x\right)\mathsf{d}x\equiv\int_{\Omega}1\mathsf{d}\rho\left(x\right),\label{EQ2}
\end{align}
from Eq. (\ref{EQ1}). This form can be seen in Eq. (\ref{EQ7}).

 \emph{Background knowledge}.---To offer a comprehensible discussion, we begin with reformulating transfer entropy, information flow, and their variants. 
 
  \begin{figure}[b!]
\includegraphics[width=1\columnwidth]{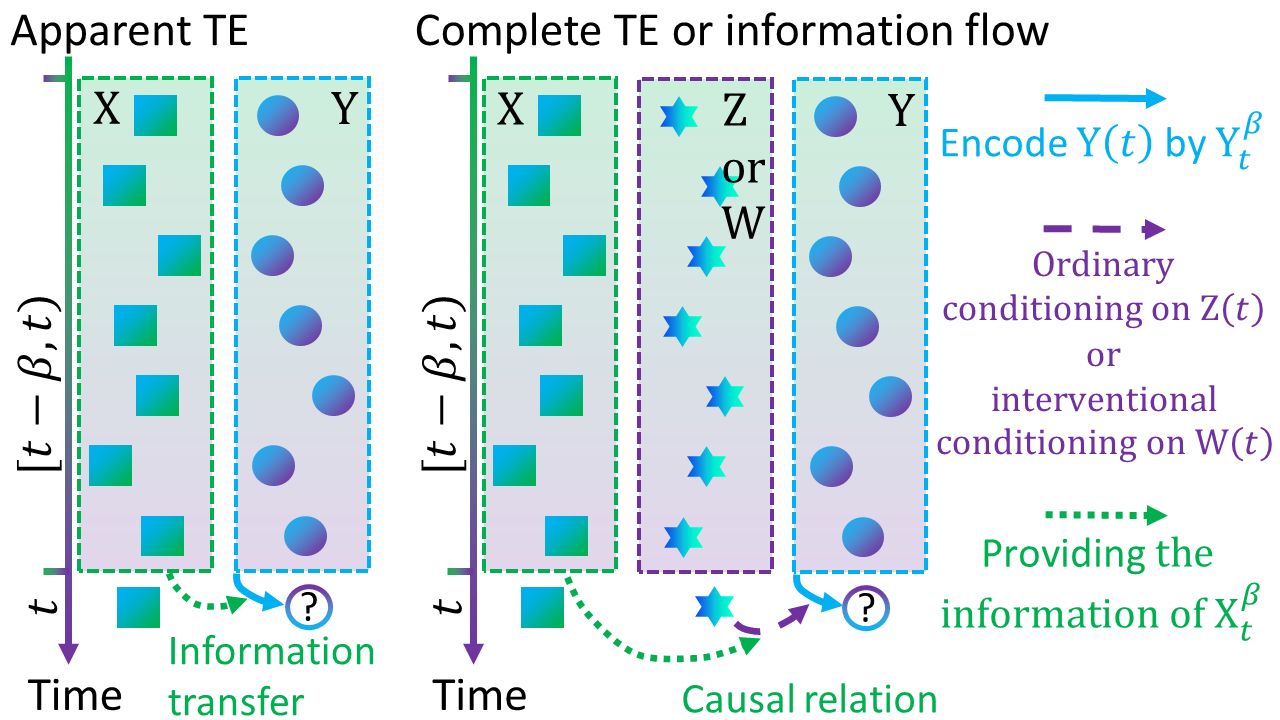}
\captionsetup{justification=raggedright}
\caption{\label{G1} The illustrations of apparent transfer entropy (TE), complete transfer entropy, and information flow.}
 \end{figure}
 
 For convenience, we consider the information transfer from a $m$-dimensional system $\mathsf{X}=\left(X_{1},\ldots,X_{m}\right)$ to a $n$-dimensional system $\mathsf{Y}=\left(Y_{1},\ldots,Y_{n}\right)$. In our derivations, we mark $\mathsf{X}_{t}^{\beta}$ as the history of $\mathsf{X}$ with a maximum time lag $\beta$, representing the historical information of $\mathsf{X}$ during a time interval $\left[t-\beta,t\right)$.
 
 The initial version of transfer entropy, referred to as \emph{apparent transfer entropy} \cite{lizier2008local,lizier2010differentiating}, is defined as \cite{schreiber2000measuring} (see Fig. \ref{G1} for illustrations)
\begin{align}
\mathcal{T}\left(\mathsf{X},\mathsf{Y}\right)=\int_{t\in\left(\beta,\infty\right)}\mathfrak{T}\left(\mathsf{X},\mathsf{Y},t\right)\mathsf{d}\rho\left(\mathsf{Y}_{t}^{\beta}\right),\label{EQ3}
\end{align}
where notion $\mathfrak{T}\left(\mathsf{X},\mathsf{Y},t\right)$ is referred to as \emph{local apparent transfer entropy} \cite{lizier2008local,lizier2010differentiating}
 \begin{align}
&\mathfrak{T}\left(\mathsf{X},\mathsf{Y},t\right)=\int_{\Omega_{\mathfrak{T}}}\Big[\log\rho\left(\mathsf{Y}\left(t\right),\mathsf{X}_{t}^{\beta}\mid\mathsf{Y}_{t}^{\beta}\right)-\notag\\&\log\rho\left(\mathsf{Y}\left(t\right)\mid\mathsf{Y}_{t}^{\beta}\right)\rho\left(\mathsf{X}_{t}^{\beta}\mid\mathsf{Y}_{t}^{\beta}\right)\Big]\mathsf{d}\rho\left(\mathsf{Y}\left(t\right),\mathsf{X}_{t}^{\beta}\mid\mathsf{Y}_{t}^{\beta}\right),\label{EQ4}
\end{align}
and $\Omega_{\mathfrak{T}}$ denotes the associated support set of probability density $\rho\left(\mathsf{Y}\left(t\right),\mathsf{X}_{t}^{\beta}\mid\mathsf{Y}_{t}^{\beta}\right)$ \cite{lizier2008local,lizier2010differentiating}. Although Eqs. (\ref{EQ3}-\ref{EQ4}) are widely used in computational works \cite{tian2021fourier,staniek2008symbolic,lobier2014phase,wollstadt2014efficient,restrepo2020transfer,zhang2019itene,kraskov2004estimating,kugiumtzis2013partial}, apparent transfer entropy actually measures the conditional correlation between $\mathsf{Y}\left(t\right)$ and $\mathsf{X}_{t}^{\beta}$ a in causal channel established by $\mathsf{Y}_{t}^{\beta}$ rather than strict causal relation \cite{lizier2008local,lizier2010differentiating}. Detailed reasons have been thoroughly analyzed by  Lizier \emph{et al.} \cite{lizier2010differentiating}, here we no longer discuss repeatedly. To distinguish causal relation from correlation, Lizier \emph{et al.} have developed \emph{complete transfer entropy} by including a selected reference system $\mathsf{Z}=\left(Z_{1},\ldots,Z_{h}\right)$ \cite{lizier2008local,lizier2010differentiating} (see Fig. \ref{G1})
\begin{align}
\mathcal{T}_{c}\left(\mathsf{X},\mathsf{Y}\mid\mathsf{Z}\right)=\int_{t\in\left(\beta,\infty\right)}\mathfrak{T}_{c}\left(\mathsf{X},\mathsf{Y}\mid\mathsf{Z},t\right)\mathsf{d}\rho\left(\mathsf{Y}_{t}^{\beta}\right),\label{EQ5}
\end{align}
 where \emph{local complete transfer entropy} is defined as 
 \begin{align}
&\mathfrak{T}_{c}\left(\mathsf{X},\mathsf{Y}\mid\mathsf{Z},t\right)=\int_{\Omega_{\mathfrak{T}_{c}}}\Bigg\{\log\rho\left(\mathsf{Y}\left(t\right),\mathsf{X}_{t}^{\beta}\mid\mathsf{Y}_{t}^{\beta},\mathsf{Z}\left(t\right)\right)-\notag\\&\log\left[\rho\left(\mathsf{Y}\left(t\right)\mid\mathsf{Y}_{t}^{\beta},\mathsf{Z}\left(t\right)\right)\rho\left(\mathsf{X}_{t}^{\beta}\mid\mathsf{Y}_{t}^{\beta},\mathsf{Z}\left(t\right)\right)\right]\Bigg\}\notag\\&\mathsf{d}\rho\left(\mathsf{Y}\left(t\right),\mathsf{X}_{t}^{\beta}\mid\mathsf{Y}_{t}^{\beta},\mathsf{Z}\left(t\right)\right),\label{EQ6}
\end{align}
and $\Omega_{\mathfrak{T}_{c}}$ is the support set of probability density $\rho\left(\mathsf{Y}\left(t\right),\mathsf{X}_{t}^{\beta}\mid\mathsf{Y}_{t}^{\beta},\mathsf{Z}\left(t\right)\right)$ \cite{lizier2008local,lizier2010differentiating}. Reference system $\mathsf{Z}$ is selected from potential information sources of $\mathsf{Y}$ in addition to $\mathsf{X}$. The optimal selection of $\mathsf{Z}$ is a critical challenge in practice and has been previously explored in Refs. \cite{verdes2005assessing,tung2007inferring,lizier2010differentiating}. 

Then we turn to reformulate \emph{information flow} \cite{ay2008information}. We introduce a notion $\dashv$ to define the action of \emph{imposing}. Imposing creates interventional conditioning by setting the value of the imposed variable. For two correlated variables $a$ and $b$, the difference between the interventional conditioning $\rho\left(b\dashv a\right)$ used for causal inference and the ordinary conditioning $\rho\left(b\mid a\right)$ used as conditional probability lies in the capacity to distinguish causal relation from correlation. Specifically, we assume that $a$ and $b$ are correlated because they are independently caused by variable $c$. For ordinary conditioning, we know $\rho\left(b\mid a\right)\neq \rho\left(b\right)$ due to the existence of correlation. For interventional conditioning, however, we have $\rho\left(b\dashv a\right)=\rho\left(b\right)$ because imposing $a$ has no influence on $b$. Based on these definitions, information flow is 
\begin{align}
&\mathcal{F}\left(\mathsf{X},\mathsf{Y}\dashv\mathsf{W}\right)=\int_{\Omega}\int_{\Lambda}\int_{\Gamma}\Bigg\{\log\rho\left(\mathsf{Y}\left(t\right)\dashv\mathsf{X}_{t}^{\beta},\mathsf{W}\left(t\right)\right)-\notag\\&\log\int_{\Upsilon}1\mathsf{d}\left[\rho\left(\mathsf{X}_{t}^{\beta}\dashv\mathsf{W}\left(t\right)\right)\rho\left(\mathsf{Y}\left(t\right)\dashv\mathsf{X}_{t}^{\beta},\mathsf{W}\left(t\right)\right)\right]\Bigg\}\notag\\&\mathsf{d}\rho\left(\mathsf{Y}\left(t\right)\dashv\mathsf{X}_{t}^{\beta},\mathsf{W}\left(t\right)\right)\mathsf{d}\rho\left(\mathsf{X}_{t}^{\beta}\dashv\mathsf{W}\left(t\right)\right)\mathsf{d}\rho\left(\mathsf{W}\left(t\right)\right),\label{EQ7}
\end{align}
where $\mathsf{W}=\left(W_{1},\ldots,W_{q}\right)$ is an associated system of the imposed variable and $t\in\left(\beta,\infty\right)$. Notions $\Omega$, $\Lambda$, $\Gamma$, and $\Upsilon$ are corresponding support sets \cite{ay2008information,lizier2010differentiating}. Please see Fig. \ref{G1} for illustrations. Although the expression of Eq. (\ref{EQ7}) seems to be complicated, its meaning is rather simple. In the terminology of causal inference, $\mathsf{X}$ is referred to a source and $\mathsf{Y}$ is a destination \cite{pearl2000models}. If $\mathsf{Y}$ is causally independent of $\mathsf{X}$, then we have 
\begin{align}
&\rho\left(\mathsf{Y}\left(t\right)\dashv\mathsf{X}_{t}^{\beta},\mathsf{W}\left(t\right)\right)\notag\\=&\rho\left(\mathsf{Y}\left(t\right)\dashv\mathsf{W}\left(t\right)\right),\label{EQ8}\\=&\int_{\Upsilon}1\mathsf{d}\left[\rho\left(\mathsf{X}_{t}^{\beta}\dashv\mathsf{W}\left(t\right)\right)\rho\left(\mathsf{Y}\left(t\right)\dashv\mathsf{X}_{t}^{\beta},\mathsf{W}\left(t\right)\right)\right],\label{EQ9}
\end{align}
meaning that observing $\mathsf{X}_{t}^{\beta}$ after having observed $\mathsf{W}\left(t\right)$ does not affect the expectation of observing $\mathsf{Y}\left(t\right)$ \cite{ay2008information}. Then, we know that Eq. (\ref{EQ7}) actually defines information flow as the deviation of destination $\mathsf{Y}$ from causal independence of source $\mathsf{X}$ after imposing a reference $\mathsf{W}$ \cite{ay2008information,lizier2010differentiating}. One can readily find that Eq. (\ref{EQ7}) has a similar form with Eqs. (\ref{EQ3}-\ref{EQ4}) and Eqs. (\ref{EQ5}-\ref{EQ6}) if we define $\mathsf{W}\left(t\right)=\mathsf{Y}_{t}^{\beta}$ (e.g., let $\mathsf{W}$ be a system of historical information of $\mathsf{Y}$). According to Ref. \cite{lizier2010differentiating}, complete transfer entropy in Eqs. (\ref{EQ5}-\ref{EQ6}) can approximate information flow when an optimal reference system $\mathsf{Z}$ is selected. Here $\mathsf{Z}$ is chosen to ensure that interventional conditioning in Eq. (\ref{EQ7}) can be expressed by  ordinary conditioning in Eqs. (\ref{EQ5}-\ref{EQ6}). One can no longer differentiate between correlation and causal relation without the reference (e.g., see Eqs. (\ref{EQ3}-\ref{EQ4}) for comparisons) \cite{lizier2010differentiating}.

 \emph{A unified view of transfer entropy and information flow}.---To this point, we have reviewed key concepts in information transfer analysis. One may notice that we reformulate them in a Kullback–Leibler divergence form of conditional mutual information \cite{cover1999elements} rather than in their original forms. The reformulation helps offer a clear version on the connections among transfer entropy, information flow, and conditional mutual information
\begin{align}
\mathcal{T}\left(\mathsf{X},\mathsf{Y}\right)&=\mathcal{I}_{\rho\left(\mathsf{X},\mathsf{Y}\right)}\left(\mathsf{X}_{t}^{\beta};\mathsf{Y}\left(t\right)\mid\mathsf{Y}_{t}^{\beta}\right),\label{EQ10}\\
\mathcal{T}_{c}\left(\mathsf{X},\mathsf{Y}\mid\mathsf{Z}\right)&=\mathcal{I}_{\rho\left(\mathsf{X},\mathsf{Y},\mathsf{Z}\right)}\left(\mathsf{X}_{t}^{\beta};\mathsf{Y}\left(t\right)\mid\mathsf{Y}_{t}^{\beta},\mathsf{Z}\left(t\right)\right),\label{EQ11}\\
\mathcal{F}\left(\mathsf{X},\mathsf{Y}\dashv\mathsf{W}\right)&=\mathcal{I}_{\rho_{\dashv}\left(\mathsf{X},\mathsf{Y},\mathsf{W}\right)}\left(\mathsf{X}_{t}^{\beta};\mathsf{Y}\left(t\right)\mid\mathsf{Y}_{t}^{\beta},\mathsf{W}\left(t\right)\right),\label{EQ12}
\end{align}
where $\rho\left(\mathsf{X},\mathsf{Y}\right)$ and $\rho\left(\mathsf{X},\mathsf{Y},\mathsf{Z}\right)$ are ordinary joint probability densities while $\rho_{\dashv}\left(\mathsf{X},\mathsf{Y},\mathsf{W}\right)$ is specially defined according to Ref. \cite{ay2008information},
\begin{align}
\rho_{\dashv}\left(\mathsf{X},\mathsf{Y},\mathsf{W}\right)=\rho\left(\mathsf{W}\right)\rho\left(\mathsf{X}\dashv\mathsf{W}\right)\rho\left(\mathsf{Y}\dashv\mathsf{X},\mathsf{W}\right).\label{EQ13}
\end{align}

These connections naturally inspire us to relate information transfer with high-order mutual information (or referred to as interaction information) \cite{timme2014synergy,schneidman2003network,ball2017multivariate,yeung1991new}. Let us take the $3$-order mutual information $\mathcal{I}_{\rho\left(\mathsf{X},\mathsf{Y},\mathsf{Z}\right)}\left(\mathsf{X};\mathsf{Y};\mathsf{Z}\right)$ between random variables $\mathsf{X}$, $\mathsf{Y}$, and $\mathsf{Z}$ as an instance. We have 
\begin{align}
&\mathcal{I}_{\rho\left(\mathsf{X},\mathsf{Y},\mathsf{Z}\right)}\left(\mathsf{X};\mathsf{Y};\mathsf{Z}\right)=\mathcal{I}_{\rho\left[\left(\mathsf{X},\mathsf{Y}\right),\mathsf{Z}\right]}\left(\mathsf{X},\mathsf{Y};\mathsf{Z}\right)-\mathcal{I}_{\rho\left(\mathsf{X},\mathsf{Z}\right)}\left(\mathsf{X};\mathsf{Z}\right)\notag\\&-\mathcal{I}_{\rho\left(\mathsf{Y},\mathsf{Z}\right)}\left(\mathsf{Y};\mathsf{Z}\right).\label{EQ14}
\end{align}
Meanwhile, we know following equations hold
\begin{align}
&\mathcal{I}_{\rho\left(\mathsf{X},\mathsf{Y},\mathsf{Z}\right)}\left(\mathsf{X};\mathsf{Y};\mathsf{Z}\right)\notag\\=&\mathcal{I}_{\rho\left(\mathsf{X},\mathsf{Y},\mathsf{Z}\right)}\left(\mathsf{X};\mathsf{Y}\mid\mathsf{Z}\right)-\mathcal{I}_{\rho\left(\mathsf{X},\mathsf{Y}\right)}\left(\mathsf{X};\mathsf{Y}\right),\label{EQ15}\\=&\mathcal{I}_{\rho\left(\mathsf{X},\mathsf{Y},\mathsf{Z}\right)}\left(\mathsf{X};\mathsf{Z}\mid\mathsf{Y}\right)-\mathcal{I}_{\rho\left(\mathsf{X},\mathsf{Z}\right)}\left(\mathsf{X};\mathsf{Z}\right),\label{EQ16}\\=&\mathcal{I}_{\rho\left(\mathsf{X},\mathsf{Y},\mathsf{Z}\right)}\left(\mathsf{Y};\mathsf{Z}\mid\mathsf{X}\right)-\mathcal{I}_{\rho\left(\mathsf{Y},\mathsf{Z}\right)}\left(\mathsf{Y};\mathsf{Z}\right).\label{EQ17}
\end{align}
Note that $\mathcal{I}_{\rho\left(\mathsf{X},\mathsf{Y},\mathsf{Z}\right)}\left(\mathsf{X};\mathsf{Y};\mathsf{Z}\right)$ in Eq. (\ref{EQ14}) can not be understood as a specific correlation among three variables where one can change the order of $\mathsf{X}$, $\mathsf{Y}$, and $\mathsf{Z}$ arbitrarily \cite{schneidman2003network}. On the contrary, $\mathcal{I}_{\rho\left(\mathsf{X},\mathsf{Y},\mathsf{Z}\right)}\left(\mathsf{X};\mathsf{Y};\mathsf{Z}\right)$ measures the difference between $\mathcal{I}_{\rho\left(\mathsf{X},\mathsf{Y},\mathsf{Z}\right)}\left(\mathsf{X};\mathsf{Y}\mid\mathsf{Z}\right)$, \emph{the information of $\mathsf{Z}$ encoded by $\mathsf{X}$ and $\mathsf{Y}$ cooperatively}, and $\mathcal{I}_{\rho\left(\mathsf{X},\mathsf{Z}\right)}\left(\mathsf{X};\mathsf{Z}\right)+\mathcal{I}_{\rho\left(\mathsf{Y},\mathsf{Z}\right)}\left(\mathsf{Y};\mathsf{Z}\right)$, \emph{the total information of $\mathsf{Z}$ encoded by $\mathsf{X}$ and $\mathsf{Y}$ independently}. There exists information synergy when $\mathcal{I}_{\rho\left(\mathsf{X},\mathsf{Y},\mathsf{Z}\right)}\left(\mathsf{X};\mathsf{Y};\mathsf{Z}\right)>0$, namely that $\mathsf{X}$ and $\mathsf{Y}$ cooperatively encode more information of $\mathsf{Z}$ than the situations where they work independently \cite{timme2014synergy,schneidman2003network,ball2017multivariate,yeung1991new}. As an opposite case, there exists information redundancy if $\mathcal{I}_{\rho\left(\mathsf{X},\mathsf{Y},\mathsf{Z}\right)}\left(\mathsf{X};\mathsf{Y};\mathsf{Z}\right)\leq 0$ \cite{timme2014synergy,schneidman2003network,ball2017multivariate,yeung1991new}.

Based on these definitions, we can naturally unify transfer entropy, information flow, and their variants based on information synergy and redundancy
\begin{align}
&\mathcal{T}\left(\mathsf{X},\mathsf{Y}\right)=\mathcal{I}_{\rho\left(\mathsf{X},\mathsf{Y}\right)}\left(\mathsf{X}_{t}^{\beta};\mathsf{Y}\left(t\right);\mathsf{Y}_{t}^{\beta}\right)\notag\\&+\mathcal{I}_{\rho\left(\mathsf{X},\mathsf{Y}\right)}\left(\mathsf{X}_{t}^{\beta};\mathsf{Y}\left(t\right)\right),\label{EQ18}\\
&\mathcal{T}_{c}\left(\mathsf{X},\mathsf{Y}\mid\mathsf{Z}\right)=\mathcal{I}_{\rho\left(\mathsf{X},\mathsf{Y},\mathsf{Z}\right)}\left(\mathsf{X}_{t}^{\beta};\mathsf{Y}\left(t\right);\mathsf{Y}_{t}^{\beta},\mathsf{Z}\left(t\right)\right)\notag\\&+\mathcal{I}_{\rho\left(\mathsf{X},\mathsf{Y}\right)}\left(\mathsf{X}_{t}^{\beta};\mathsf{Y}\left(t\right)\right),\label{EQ19}\\
&\mathcal{F}\left(\mathsf{X},\mathsf{Y}\dashv\mathsf{W}\right)=\mathcal{I}_{\rho_{\dashv}\left(\mathsf{X},\mathsf{Y},\mathsf{W}\right)}\left(\mathsf{X}_{t}^{\beta};\mathsf{Y}\left(t\right);\mathsf{Y}_{t}^{\beta},\mathsf{W}\left(t\right)\right)\notag\\&+\mathcal{I}_{\rho_{\dashv}\left(\mathsf{X},\mathsf{Y}\right)}\left(\mathsf{X}_{t}^{\beta};\mathsf{Y}\left(t\right)\right).\label{EQ20}
\end{align}
Eqs. (\ref{EQ18}-\ref{EQ20}) present plentiful interesting insights on our concerned questions (I-III). Below, we discuss each of the question. 

\emph{How does information transfer arise and vanish}.---In Eq. (\ref{EQ18}), we can see that there exists ordinary information transfer, irrespective of being causal relation or not, if and only if
\begin{align}
    \mathcal{I}_{\rho\left(\mathsf{X},\mathsf{Y}\right)}\left(\mathsf{X}_{t}^{\beta};\mathsf{Y}\left(t\right);\mathsf{Y}_{t}^{\beta}\right)+\mathcal{I}_{\rho\left(\mathsf{X},\mathsf{Y}\right)}\left(\mathsf{X}_{t}^{\beta};\mathsf{Y}\left(t\right)\right)\neq 0.\label{EQ21}
\end{align}

If we reformulate the condition for ordinary information transfer to vanish, the opposite case of Eq. (\ref{EQ21}), based on Eq. (\ref{EQ14}), we derive
\begin{align}
   &\mathcal{I}_{\rho\left(\mathsf{X},\mathsf{Y}\right)}\left(\mathsf{X}_{t}^{\beta},\mathsf{Y}\left(t\right);\mathsf{Y}_{t}^{\beta}\right)=-\mathcal{I}_{\rho\left(\mathsf{X},\mathsf{Y}\right)}\left(\mathsf{X}_{t}^{\beta};\mathsf{Y}\left(t\right)\right)+\notag\\&\mathcal{I}_{\rho\left(\mathsf{X},\mathsf{Y}\right)}\left(\mathsf{X}_{t}^{\beta};\mathsf{Y}_{t}^{\beta}\right)+\mathcal{I}_{\rho\left(\mathsf{Y}\right)}\left(\mathsf{Y}\left(t\right);\mathsf{Y}_{t}^{\beta}\right),\label{EQ22}
\end{align}
Understanding Eq. (\ref{EQ22}) from the perspective of causal inference \cite{pearl2000models} is rather simple. Because of the symmetry of mutual information \cite{cover1999elements}, we can equivalently interpret terms $\mathcal{I}_{\rho\left(\mathsf{X},\mathsf{Y}\right)}\left(\mathsf{X}_{t}^{\beta},\mathsf{Y}\left(t\right);\mathsf{Y}_{t}^{\beta}\right)$, $\mathcal{I}_{\rho\left(\mathsf{X},\mathsf{Y}\right)}\left(\mathsf{X}_{t}^{\beta};\mathsf{Y}_{t}^{\beta}\right)$, and $\mathcal{I}_{\rho\left(\mathsf{Y}\right)}\left(\mathsf{Y}\left(t\right);\mathsf{Y}_{t}^{\beta}\right)$ in Eq. (\ref{EQ22}) as the encoded information of $\left(\mathsf{X}_{t}^{\beta},\mathsf{Y}\left(t\right)\right)$, $\mathsf{X}_{t}^{\beta}$, and $\mathsf{Y}\left(t\right)$ by $\mathsf{Y}_{t}^{\beta}$, respectively. Meanwhile, we can understand $\mathcal{I}_{\rho\left(\mathsf{X},\mathsf{Y}\right)}\left(\mathsf{X}_{t}^{\beta};\mathsf{Y}\left(t\right)\right)$ as the shared part between $\mathsf{X}_{t}^{\beta}$ and $\mathsf{Y}\left(t\right)$. Combining these interpretations, we can see that Eq. (\ref{EQ22}) actually means that the encoded information of $\left(\mathsf{X}_{t}^{\beta},\mathsf{Y}\left(t\right)\right)$ by $\mathsf{Y}_{t}^{\beta}$ excludes the coupled part of $\mathsf{X}_{t}^{\beta}$ and $\mathsf{Y}\left(t\right)$, making these two variables independent from each other in the view of $\mathsf{Y}_{t}^{\beta}$. In other words, the current state of system $\mathsf{Y}$ is independent of the historical information of system $\mathsf{X}$ given the historical information of system $\mathsf{Y}$ 
\begin{align}
   \mathsf{Y}\left(t\right)\perp\!\!\!\perp\mathsf{X}_{t}^{\beta}\mid\mathsf{Y}_{t}^{\beta},\;\forall t\in\left(\beta,\infty\right),\label{EQ23}
\end{align}
where notion $\perp\!\!\!\perp$ denotes independence. Eqs. (\ref{EQ22}-\ref{EQ23}) means that ordinary information transfer is a special case of information synergy and redundancy where Eq. (\ref{EQ22}) is unsatisfied. 

We can derive a non-trivial finding by reformulating Eq. (\ref{EQ22}) with the chain rule of mutual information \cite{cover1999elements}
\begin{align}
   &\mathcal{I}_{\rho\left(\mathsf{X},\mathsf{Y}\right)}\left(\mathsf{Y}_{t}^{\beta};\mathsf{Y}\left(t\right)\mid\mathsf{X}_{t}^{\beta}\right)=\mathcal{I}_{\rho\left(\mathsf{Y}\right)}\left(\mathsf{Y}\left(t\right);\mathsf{Y}_{t}^{\beta}\right)\notag\\&-\mathcal{I}_{\rho\left(\mathsf{X},\mathsf{Y}\right)}\left(\mathsf{X}_{t}^{\beta};\mathsf{Y}\left(t\right)\right).\label{EQ24}
\end{align}
 Eq. (\ref{EQ24}) means that ordinary information transfer vanishes because adding the historical information $\mathsf{X}_{t}^{\beta}$ of $\mathsf{X}$ does not supply more information for $\mathsf{Y}$ to predict the target state $\mathsf{Y}\left(t\right)$ based on $\mathsf{Y}_{t}^{\beta}$. On the contrary, providing $\mathsf{X}_{t}^{\beta}$ reduces the original encoded information of $\mathsf{Y}\left(t\right)$ in $\mathsf{Y}_{t}^{\beta}$ by a quantity of $\mathcal{I}_{\rho\left(\mathsf{X},\mathsf{Y}\right)}\left(\mathsf{X}_{t}^{\beta};\mathsf{Y}\left(t\right)\right)$. We refer to this phenomenon as \emph{information damage}, where introducing $d$-bits historical information of $\mathsf{X}$ will damage the encoding process of the target state of $\mathsf{Y}$ by $d$-bits. Ordinary information transfer occurs if and only if there is no information damage. Please note that conditional mutual information is always non-negative. Therefore, there always exists information transfer if 
 \begin{align}
   \mathcal{I}_{\rho\left(\mathsf{Y}\right)}\left(\mathsf{Y}\left(t\right);\mathsf{Y}_{t}^{\beta}\right)<\mathcal{I}_{\rho\left(\mathsf{X},\mathsf{Y}\right)}\left(\mathsf{X}_{t}^{\beta};\mathsf{Y}\left(t\right)\right).\label{EQ25}
\end{align}
 Eq. (\ref{EQ25}) describes an interesting property where information transfer from $\mathsf{X}$ to $\mathsf{Y}$, just like a kind of diffusion, occurs if there is an abstract information gradient from the encoded information of $\mathsf{Y}\left(t\right)$ in $\mathsf{X}_{t}^{\beta}$ (higher) to the encoded information of $\mathsf{Y}\left(t\right)$ in $\mathsf{Y}_{t}^{\beta}$ (lower). We refer to the case in Eq. (\ref{EQ25}) as \emph{natural information transfer} while its opposite case is called \emph{conditional information transfer}. Moreover, we can define a measure of information transfer effect size. For natural information transfer, we define its effect size as
 \begin{align}
  \mathcal{T}^{\text{effect}}\left(\mathsf{X},\mathsf{Y}\right)=\log\frac{\mathcal{I}_{\rho\left(\mathsf{X},\mathsf{Y}\right)}\left(\mathsf{Y}\left(t\right);\mathsf{X}_{t}^{\beta}\right)}{\mathcal{I}_{\rho\left(\mathsf{Y}\right)}\left(\mathsf{Y}\left(t\right);\mathsf{Y}_{t}^{\beta}\right)}.\label{EQ26}
\end{align}
For conditional information transfer, we measure the deviation of the actual quantity of $\mathcal{I}_{\rho\left(\mathsf{X},\mathsf{Y}\right)}\left(\mathsf{Y}_{t}^{\beta};\mathsf{Y}\left(t\right)\mid\mathsf{X}_{t}^{\beta}\right)$ from information damage as its effect size
 \begin{align}
  &\mathcal{T}^{\text{effect}}\left(\mathsf{X},\mathsf{Y}\right)\notag\\=&\log\frac{\mathcal{I}_{\rho\left(\mathsf{X},\mathsf{Y}\right)}\left(\mathsf{Y}_{t}^{\beta};\mathsf{Y}\left(t\right)\mid\mathsf{X}_{t}^{\beta}\right)}{\mathcal{I}_{\rho\left(\mathsf{Y}\right)}\left(\mathsf{Y}\left(t\right);\mathsf{Y}_{t}^{\beta}\right)-\mathcal{I}_{\rho\left(\mathsf{X},\mathsf{Y}\right)}\left(\mathsf{X}_{t}^{\beta};\mathsf{Y}\left(t\right)\right)}.\label{EQ27}
\end{align}

\emph{How does causal relation arise, vanish, and differ from information transfer}.---Similar analyses can also be implemented when we turn to analyzing causal relation in Eqs. (\ref{EQ19}-\ref{EQ20}). Following analogous derivations in Eqs. (\ref{EQ21}-\ref{EQ25}), we can readily know 
\begin{align}
   &\mathsf{Y}\left(t\right)\perp\!\!\!\perp\mathsf{X}_{t}^{\beta}\mid\left(\mathsf{Y}_{t}^{\beta},\mathsf{Z}\left(t\right)\right),\;\forall t\in\left(\beta,\infty\right),\label{EQ28}
\end{align}
if and only if 
\begin{align}
    &\mathcal{I}_{\rho\left(\mathsf{X},\mathsf{Y}\right)}\left(\mathsf{Y}_{t}^{\beta};\mathsf{Y}\left(t\right)\mid\mathsf{X}_{t}^{\beta}\right)=\mathcal{I}_{\rho\left(\mathsf{Y},\mathsf{Z}\right)}\left(\mathsf{Y}\left(t\right);\mathsf{Y}_{t}^{\beta}\mid\mathsf{Z}\left(t\right)\right)\notag\\&-\mathcal{I}_{\rho\left(\mathsf{X},\mathsf{Y}\right)}\left(\mathsf{X}_{t}^{\beta};\mathsf{Y}\left(t\right)\right)+\mathcal{I}_{\rho\left(\mathsf{Y},\mathsf{Z}\right)}\left(\mathsf{Z}\left(t\right);\mathsf{Y}\left(t\right)\right)\notag\\&-\mathcal{I}_{\rho\left(\mathsf{X},\mathsf{Y},\mathsf{Z}\right)}\left(\mathsf{Z}\left(t\right);\mathsf{Y}\left(t\right)\mid\mathsf{Y}_{t}^{\beta},\mathsf{X}_{t}^{\beta}\right).\label{EQ29}
\end{align}
Meanwhile, we have 
\begin{align}
   &\mathsf{Y}\left(t\right)\perp\!\!\!\perp\mathsf{X}_{t}^{\beta}\dashv\left(\mathsf{Y}_{t}^{\beta},\mathsf{W}\left(t\right)\right),\;\forall t\in\left(\beta,\infty\right),\label{EQ30}
\end{align}
if and only if 
\begin{align}
    &\mathcal{I}_{\rho_{\dashv}\left(\mathsf{X},\mathsf{Y}\right)}\left(\mathsf{Y}_{t}^{\beta};\mathsf{Y}\left(t\right)\mid\mathsf{X}_{t}^{\beta}\right)=\mathcal{I}_{\rho_{\dashv}\left(\mathsf{Y},\mathsf{W}\right)}\left(\mathsf{Y}\left(t\right);\mathsf{Y}_{t}^{\beta}\mid\mathsf{W}\left(t\right)\right)\notag\\&-\mathcal{I}_{\rho_{\dashv}\left(\mathsf{X},\mathsf{Y}\right)}\left(\mathsf{X}_{t}^{\beta};\mathsf{Y}\left(t\right)\right)+\mathcal{I}_{\rho_{\dashv}\left(\mathsf{Y},\mathsf{W}\right)}\left(\mathsf{W}\left(t\right);\mathsf{Y}\left(t\right)\right)\notag\\&-\mathcal{I}_{\rho_{\dashv}\left(\mathsf{X},\mathsf{Y},\mathsf{W}\right)}\left(\mathsf{W}\left(t\right);\mathsf{Y}\left(t\right)\mid\mathsf{Y}_{t}^{\beta},\mathsf{X}_{t}^{\beta}\right).\label{EQ31}
\end{align}
Taking Eq. (\ref{EQ29}) as an instance, we can see that causal relation vanishes if following conditions are satisfied: (a) including the historical information $\mathsf{X}_{t}^{\beta}$ of $\mathsf{X}$ implies information damage to reduce the encoded information of $\mathsf{Y}\left(t\right)$ in $\mathsf{Y}_{t}^{\beta}$ given $\mathsf{Z}\left(t\right)$ by a quantity of $\mathcal{I}_{\rho\left(\mathsf{X},\mathsf{Y}\right)}\left(\mathsf{X}_{t}^{\beta};\mathsf{Y}\left(t\right)\right)$; (b) including $\mathsf{X}_{t}^{\beta}$ also damages the the encoded information of $\mathsf{Y}\left(t\right)$ in $\mathsf{Z}\left(t\right)$ and reduces the encoded information by a quantity of $\mathcal{I}_{\rho\left(\mathsf{X},\mathsf{Y},\mathsf{Z}\right)}\left(\mathsf{Z}\left(t\right);\mathsf{Y}\left(t\right)\mid\mathsf{Y}_{t}^{\beta},\mathsf{X}_{t}^{\beta}\right)$. While condition (a) is similar to the situation of ordinary information transfer in Eq. (\ref{EQ24}), condition (b) differentiates causal relation from ordinary information transfer. Please note that there may exist multiple choices of reference system $\mathsf{Z}$. Therefore, causal relation has higher possibility to vanish than ordinary information transfer. Eq. (\ref{EQ31}) can be understood in a similar manner so we no longer repeat. In general, Eqs. (\ref{EQ28}-\ref{EQ31}) suggest that causal relation is a special case of information synergy and redundancy phenomenon where Eq. (\ref{EQ29}) or Eq. (\ref{EQ31}) is unsatisfied. 

Moreover, we can analogously define the effect size of causal relation. We refer to the case where the right sides of Eq. (\ref{EQ29}) and Eq. (\ref{EQ31}) are negative as \emph{natural causal relation} and measure its effect size as
 \begin{align}
  &\mathcal{F}^{\text{effect}}\left(\mathsf{X},\mathsf{Y}\dashv\mathsf{W}\right)\notag\\\simeq&\mathcal{T}_{c}^{\text{effect}}\left(\mathsf{X},\mathsf{Y}\mid\mathsf{Z}\right)\label{EQ32}\\=&\log\frac{\mathcal{I}_{\rho\left(\mathsf{X},\mathsf{Y}\right)}\left(\mathsf{X}_{t}^{\beta};\mathsf{Y}\left(t\right)\right)\mathcal{I}_{\rho\left(\mathsf{X},\mathsf{Y},\mathsf{Z}\right)}\left(\mathsf{Z}\left(t\right);\mathsf{Y}\left(t\right)\mid\mathsf{Y}_{t}^{\beta},\mathsf{X}_{t}^{\beta}\right)}{\mathcal{I}_{\rho\left(\mathsf{Y},\mathsf{Z}\right)}\left(\mathsf{Y}\left(t\right);\mathsf{Y}_{t}^{\beta}\mid\mathsf{Z}\left(t\right)\right)\mathcal{I}_{\rho\left(\mathsf{Y},\mathsf{Z}\right)}\left(\mathsf{Z}\left(t\right);\mathsf{Y}\left(t\right)\right)}.\label{EQ33}
\end{align}
We refer to the case where the right sides of Eq. (\ref{EQ29}) and Eq. (\ref{EQ31}) are positive as \emph{conditional causal relation} and define 
 \begin{align}
  \mathcal{F}^{\text{effect}}\left(\mathsf{X},\mathsf{Y}\dashv\mathsf{W}\right)&\simeq\mathcal{T}_{c}^{\text{effect}}\left(\mathsf{X},\mathsf{Y}\mid\mathsf{Z}\right)\label{EQ34}\\&=\log\frac{\mathcal{I}_{\rho\left(\mathsf{X},\mathsf{Y}\right)}\left(\mathsf{Y}_{t}^{\beta};\mathsf{Y}\left(t\right)\mid\mathsf{X}_{t}^{\beta}\right)}{\chi\left(\mathsf{X},\mathsf{Y},\mathsf{Z}\right)}\label{EQ35}
\end{align}
as its causal effect size, where $\chi\left(\mathsf{X},\mathsf{Y},\mathsf{Z}\right)$ denotes the right side of Eq. (\ref{EQ29}). Note that reference sources $\mathsf{Z}$ and $\mathsf{W}$ in Eqs. (\ref{EQ32}-\ref{EQ35}) should be selected appropriately in practice such that $\mathcal{F}^{\text{effect}}\left(\mathsf{X},\mathsf{Y}\dashv\mathsf{W}\right)\simeq\mathcal{T}_{c}^{\text{effect}}\left(\mathsf{X},\mathsf{Y}\mid\mathsf{Z}\right)$ principally holds \cite{lizier2010differentiating}.

\emph{Conclusion: information transfer and causal relation are special cases of a unified and general variable}.---We have presented a unified theory to suggest that information transfer and causal relation are special cases of information synergy and redundancy that require additional conditions. We analytically show that information synergy and redundancy theory can explain the preconditions for information transfer and causal relation to exist or vanish, the difference between ordinary information transfer and causal relation, and the effect sizes of information transfer and causal relation. Compared with the well-known relation between information transfer and \emph{Granger causality} that only hold on Gaussian variables \cite{barnett2009granger}, the discovered relations in our research have no limitation on random variable properties. Although there remain plentiful detailed questions for future explorations in this preliminary study (e.g., analyzing information transfer and causal relation with the Yeung's bound on $3$-order mutual information \cite{yeung1991new}), we anticipate that our framework may serve as a potential attempt to unify Pearl’s causal inference theory in computer science \cite{pearl2000models} and information transfer analysis in physics \cite{schreiber2000measuring,ay2008information,lizier2010differentiating}. 
 
\emph{Acknowledgements}.---This project is supported by the Artificial and General Intelligence Research Program of Guo Qiang Research Institute at Tsinghua University (2020GQG1017) as well as the Tsinghua University Initiative Scientific Research Program. 


\bibliography{apssamp}
\end{document}